\documentstyle[12pt]{article}

\def\beqa{\begin{eqnarray}}
\def\eeqa{\end{eqnarray}}
\def\beq{\begin{equation}}
\def\eeq{\end{equation}}

\def\unu{^{\nu}}
\def\dmu{_{\mu}}  

\def\umunu{^{\mu\nu}}
\def\dmunu{_{\mu\nu}}

\def\ddemunu{_{;\mu\nu}}

\def\ddemu{_{;\mu}}  
\def\ddenu{_{;\nu}}  
\def\ddea{_{;\alpha}}  \def\udea{^{;\alpha}}

\def\bib#1{$^{\ref{#1}}$}

\let\gam=\gamma
\let\alp=\alpha

\def\pr{{\it Phys. Rev.}\ }
\def\prl{{\it Phys. Rev. Lett.}\ }
\def\pl{{\it Phys. Lett.}\ }
\def\np{{\it Nucl. Phys.}\ }

\def\ijmp{{\it Int. Journ. Mod. Phys.}\ }
\def\ijtp{{\it Int. Journ. Theor. Phys.}\ }

\def\cqg{{\it Class. Quantum Grav.}\ }

\def\f{F(\phi)}
\def\fp{F'(\phi)}

\def\p{\phi}

\def\v{V(\phi)}
\def\vp{V'(\phi)}
\def\l{\cal L}
\begin{document}
\def\bib#1{[{\ref{#1}}]}
\begin{titlepage}
\title{A time--dependent Cosmological "Constant"}

 \author{{S. Capozziello, R. de Ritis}
\\ {\em Dipartimento di Scienze Fisiche, Universit\`{a} di Napoli,}
\\ {\em Istituto Nazionale di Fisica Nucleare, Sezione di Napoli,}\\
   {\em Mostra d'Oltremare pad. 19 I-80125 Napoli, Italy.}}
	      \date{}
	      \maketitle
	      \begin{abstract}
We extend  the Wald cosmic no--hair theorem to a general class of 
scalar--tensor nonminimally coupled  theories of gravity where
ordinary matter is also present in the form of a perfect fluid.
We give a  set of conditions for obtaining 
a de Sitter expansion independently of any initial conditions, generalizing 
the treatment of such a problem as given by Wald. Finally we apply the 
results to some specific models.

	      \end{abstract}

\vspace{20. mm}
e-mail address:\\
 capozziello@axpna1.na.infn.it\\
 deritis@axpna1.na.infn.it
	      \vfill
	      \end{titlepage}

\section{\normalsize \bf Introduction}
In this paper we  generalize the so called 
"{\it cosmological no--hair theorem}" \bib{wald},
to the case of nonminimally coupled scalar--tensor 
theories of gravity. These ones were formulated  years ago in order to 
better understand
inertia and Mach's Principle in the theoretical framework 
of General Relativity. Today, they are playing an important 
role in cosmology, 
due to the fact that they seem to solve a lot of shortcomings connected 
with inflationary models like those related to the phase transition from 
false vacuum to true vacuum in early universe and those related to the bubble 
spectrum  able to produce seeds for the large scale structures
(see \bib{bergmann},\bib{brans},\bib{cimento} and 
references therein for a discussion of nonminimally 
coupled scalar-tensor theories).

The no--hair conjecture was introduced by Hoyle and Narlikar \bib{hoyle}:
 roughly speaking, they claimed that if there is a positive cosmological 
constant, all
the expanding universes will approach  the de Sitter behaviour. 
That is, if a cosmological constant is present, 
no matter which are the initial 
conditions, the universe will become  homogeneous and isotropic.
However, there is something  vague in such a formulation. It is not clear 
the meaning of {\it "expanding universe"} as well as 
{\it "approach the de Sitter behaviour"}. 
Furthermore, there is no general proof (or disproof) of such a conjecture;
on the contrary, there are counter--examples of initially expanding and then
recollapsing universes which never become de Sitter \bib{cotsakis}.

In 1983, Wald  gave a proof of a simplified version of the conjecture.
He proved that:

 {\it All Bianchi cosmologies (except IX), in presence of 
a positive cosmological constant, asymptotically approach the de Sitter 
behaviour} \bib{wald}. 

In all these discussions (in  Wald's paper too),
the cosmological constant is a true constant and it is put by hand in the 
gravitational arena. 
It is relevant that in Wald's proof are not used the Bianchi 
identities, then the  proof is independent of the
dynamical evolution of the material sources. 
Here we discuss how to introduce a time dependent 
"cosmological constant" in the 
 context of scalar--tensor theories of gravity in which a scalar field $\p$
is nonminimally coupled to geometry being also present a standard perfect 
fluid matter (noninteracting with $\p$),(see also \bib{Pimentel}) . 
So doing, we will introduce a more 
general set of conditions respect to those given by Wald,
not based on various "energy conditions" as in Wald, 
by which it is possible to get a {\it de Sitter asymptotic}
behaviour for the scale factor of the universe;
in other words, we introduce an {\it "effective
cosmological constant"} via the asymptotic de Sitter expansion.
Furthermore, we will show that the de Sitter asymptotic behaviour 
is not depending 
on the (asymptotic) sign, as well as on the value, of the effective  
gravitational  "constant", which is, in our units $(8\pi G_{N}=\hbar=c=1)$,
\beq
\label{newton}
G_{eff}=-\frac{1}{2 F(\p)}\,,
\eeq
where $\f$ describes the (nonminimal) gravitational coupling.
In what follows, we take into account only  
Friedman--Robertson--Walker (FRW) flat cosmologies 
described by the scale factor $a(t)$. 
The scalar field $\p$ is, of course, a function of time only.

The paper is organized as follows: we discuss first how to introduce 
a cosmological constant, then we generalize  Wald's theorem. Finally we 
give some concrete  examples.

\section{\normalsize \bf The Effective Cosmological Constant}

The class of theories we are considering are described by the action
\beq
\label{1}
{\cal A}=\int d^{4}x\sqrt{-g}\left[\f R +\frac{1}{2}
g\umunu \phi\ddemu \phi\ddenu-V(\phi)+{\l}_{m}\right],
\eeq
where $\v$ and $\f$  are generic functions of  the field $\p$ and
${\l}_{m}$ is the ordinary matter contribution to the total
Lagrangian density.

The Einstein equations can be written as 
\beq
\label{2}
G\dmunu=\tilde{T}\dmunu=-\frac{1}{2\f}T^{(tot)}\dmunu\,,
\eeq
where $G\dmunu$ is the Einstein tensor and $T^{(tot)}\dmunu$ is the total
stress--energy tensor
\beq
\label{3}
T^{(tot)}\dmunu=T^{(\p)}\dmunu+T^{(m)}\dmunu
\eeq
where
\beq
\label{4}
T^{(\p)}\dmunu=
\phi\ddemu \phi\ddenu-\frac{1}{2} g\dmunu \phi\ddea \phi\udea+g\dmunu\v
+2g\dmunu\Box\f-2\f\ddemunu\,,
\eeq
represents the scalar field source and $T^{(m)}\dmunu$ is the perfect
fluid matter source.
The Klein--Gordon (KG) equation is obtained from the action (\ref{1}) 
by  varying with respect to $\p$:
\beq
\label{5}
\Box \p-R\fp+\vp=0\,,
\eeq
where the prime means the derivative with respect to $\p$. 
Using the Einstein equations,
from the contracted Bianchi identity \bib{nmc}, we obtain the relation: 
\beq
[\f G\dmu\unu -T^{(\p)}\dmu{\unu}]\ddenu = 
\frac{1}{2}\p_{;\mu}[\Box \p -R\fp +\vp]=T^{(m)}\dmu{\unu}{\ddenu}\,;
\eeq
then imposing  the KG equation, we get
$T^{(m)}\dmu{\unu}{\ddenu}=0$, which is the usual " Bianchi identity" 
for standard matter.

The major point of our discussion is the following: it is possible to
 construct a time--dependent cosmological 
"constant" coherently with the Einstein equations
as well as to the (contracted) Bianchi identity? In other words, 
for constructing an effective (time--dependent) cosmological 
"constant", we cannot refer to  
the standard stress--energy tensor of the form $\Lambda g\dmunu$ since this
implies the introduction of a truly constant $\Lambda$.

As we already said, we will  restrict our considerations 
only to FRW (flat) universes. In other words, we will
introduce an effective (time dependent) cosmological "constant"
only in a cosmological context. We will discuss this important point
in the conclusions.
In this case  the metric  is:
\beq
\label{6}
ds^{2}=dt^{2}-a(t)^{2}(dx^{2}+dy^{2}+dz^{2})\,,
\eeq
where  $a=a(t)$ is the scale factor of the universe. 
From (\ref{2}), using  metric (\ref{6}), we get
the (cosmological) Einstein equations
\beq
\label{7}
H^{2}+\frac{\dot{F}}{F}H +\frac{\rho_{\p}}{6F}+\frac{\rho_{m}}{6F}=0\,,
\eeq
\beq
\label{8}
\dot{H} = -\left(H^2 +\frac{V}{6F}\right) - H\frac{\dot{F}}{2F} +
\frac{\dot{\p}^2}{6F}-
-\frac{1}{2}\frac{\ddot{F}}{F} + \frac{3p_{m}+\rho_{m}}{12 F}.
\eeq
where $H=\dot{a}/a$, $\rho_{\p}=\frac{1}{2}\dot{\p}^{2}+\v$,
 $\rho_{m}$, $p_{m}$ are, respectively,  
the Hubble parameter, the energy density of scalar field, the energy
density and the pressure  of
standard matter. We have not used 
the state equation of standard matter.
Eq.(\ref{7}) can be rewritten as:
\beq
\label{7'}
{\cal P}(H) \equiv 
\left(H-\Lambda_{eff,\,1}\right)\left(H-\Lambda_{eff,\,2}\right)=
-\frac{\rho_{m}}{6F}\,,
\eeq
where ${\cal P}(H)$ is a second degree polinomial in $H$, and
\beq
\label{9}
\Lambda_{eff,\,1,2}=-\frac{\dot{F}}{2F}\pm
\sqrt{\left(\frac{\dot{F}}{2F}\right)^{2}-\frac{\rho_{\p}}{6F}}\,,
\eeq
"1" is relative to  choose the plus sign and "2" to the minus.
Using  
the effective gravitational coupling (\ref{newton}) 
and its time (relative) 
variation,the quantities defined by Eq.(\ref{9}) can be rewritten:
\beq
\label{10}
\Lambda_{eff,\,1,2}=\frac{\dot{G}_{eff}}{2G_{eff}}\pm
\sqrt{\left(\frac{\dot{G}_{eff}}{2G_{eff}}\right)^{2}+
\frac{G_{eff}\rho_{\p}}{3}}\,.
\eeq
Furthermore, 
$\Lambda_{eff,\,1,2}$ have to be real, then the restriction
\beq
\label{real}
\left(\frac{\dot{F}}{2F}\right)^{2}\geq\frac{\rho_{\p}}{6F}\,,
\eeq
has to be satisfied.
From the definition we have given, we get:
\beq
\Lambda_{eff,\,1}+\Lambda_{eff,\,2}=-\frac{\dot{F}}{F}\,,\;\;\;\;
\Lambda_{eff,\,1}-\Lambda_{eff,\,2}=
2\sqrt{\left(\frac{\dot{F}}{2F}\right)^{2}-\frac{\rho_{\p}}{6F}}\geq 0\,,
\eeq
that is, in general,
\beq
\label{11}
\Lambda_{eff,\,1}\geq\Lambda_{eff,\,2} \,.
\eeq
In terms of ${\cal P}(H)$, Eq.(\ref{8}) for $\dot{H}$ becomes
\beq
\label{8''}
\dot{H} +{\cal P}(H)=
H \frac{\dot{F}}{2 F}+ \frac{\dot{\p}^2}{4 F} - 
\frac{1}{2}\frac{d}{dt}\left(\frac{\dot{F}}{F}\right)-
\frac{1}{2}\left(\frac{\dot{F}}{F}\right)^2 +
\frac{3p_{m}+\rho_{m}}{12 F}\,.
\eeq
We make now the following two hypotheses on the asymptotic behaviour
of the $\f$, i.e. for $t\gg 0$, we suppose that:
\beq
\label{i}
\frac{\dot{F}}{F}\longrightarrow \Sigma_{0}\;,
\eeq
\beq
\label{ii}
\frac{\rho_{\p}}{6\f}\longrightarrow\Sigma_{1}\;;
\eeq
where $\Sigma_{0,1}$ are two constants depending,respectively,on the parameters
of the coupling and of the coupling and of the potential. 
Under these two hypotheses we see that
the two quantities $\Lambda_{eff,\,1,2}$ asymptotically go to constants.
{\it Viceversa}, if we assume that
$\Lambda_{eff,\,i}\rightarrow \Lambda_{i}$ (constants), 
we see that $\dot{F}/F$ and $\rho_{\p}/6F$ become
constants for large $t$.
Then hypotheses (\ref{i}) and (\ref{ii}) 
are necessary and sufficient conditions
since the two $\Lambda$'s are asymptotically constants.
It is important to stress that hypothesis (\ref{i}) does not select a specific 
asymptotic behaviour for $\dot{F}/F$ since a wide class of $G_{eff}$
is allowed.

We will also assume, 
in the following considerations, 
that asymptotically the sign of $\f$ is constant 
(this is our third, quite natural, assumption), 
and
then we have to consider
 the two cases: $F(t\gg 0)\leq 0$ and   $F(t\gg 0)\geq 0$.
Since  we are considering that, asymptotically,
$\dot{F}/F$ is constant,
 each of the above cases has two subcases  related
to the sign of $\dot{F}$. Of course  the case
$F(t\gg 0)\leq 0$ is physically relevant: 
the other one (repulsive gravity) can be interesting if related to the 
possibility of recovering the de Sitter behaviour
for $a(t)$. In this way it appears clear that 
recovering such an asymptotic 
behaviour for $a(t)$ in not even connected to recover the standard sign of 
gravity, as we will discuss below, in general and in 
connection to some concrete examples.
Let us  now consider the case $F(t\gg 0)\leq 0$ and $\dot{F}(t\gg 0)\leq 0$;
from hypothesis (\ref{i}) we have $\Sigma_{0}\geq 0$. Furthermore
the condition (\ref{real}) is (asymptotically) satisfied.
Eq.(\ref{7'}) gives
\beq
\label{7'''}
{\cal P}(H)\geq 0,
\eeq
then we have $H \geq \Lambda_{1}$, $H \leq \Lambda_{2}$.
For the two $\Lambda_{i}$, we obtain the asymptotic expressions:
\beq
\Lambda_{1} = -\frac{\Sigma_{0}}{2} + 
\sqrt{\left(\frac{\Sigma_{0}}{2}\right)^2 +|\Sigma_{1}|} \geq 0\,,
\eeq
\beq
\Lambda_{2} = -\frac{\Sigma_{0}}{2} -
\sqrt{\left(\frac{\Sigma_{0}}{2}\right)^2 +|\Sigma_{1}|} \leq 0\,.
\eeq
Considering Eq.(\ref{8}), 
 we have
\beq
\label{8''''}
\dot{H} = -\left(H^2 -\frac{V}{6|F|}\right) -H \frac{\dot{F}}{2F} -
\frac{\dot{\p}^2}{6|F|}-
\frac{1}{2}(\frac{\ddot{F}}{F}) - \frac{3p_{m}+\rho_{m}}{12|F|}.
\eeq
If (this is our last hypothesis)
\beq
\label{iii}
H^2 \geq \frac{V}{6|F|},
\eeq
we obtain then
\beq
\label{8'''''}
\dot{H} \leq 0\,.
\eeq
In other words, from the two disequalities on ${\cal{P}}(H)$
and on $\dot{H}$ we find that $H(t)$ has  a horizontal asymptote,
or, equivalently,  $H$ goes to a constant. 
Then the universe, for large $t$, has a de Sitter behaviour,
(i.e. $a(t)\sim \exp (\alp t)$,  where $\alp$ is an unknown constant).
Under the  conditions (\ref{i}),(\ref{ii}), the constant 
asymptotic sign of $F(\p(t))$ and under the condition
(\ref{iii}),   the 
universe, for large $t$, expands as de Sitter, even if  it is not fixed
the parameter which specifies such an expansion, i.e. 
the effective cosmological constant.         
If we compare  Wald's conditions with ours, we have:
\beqa
\mbox{(Wald)}\;\;\;\;\;\; & &\;\;\;\mbox{(our  conditions)}\nonumber\\
\left(H-\sqrt{\frac{\Lambda}{3}}\right)\left(H
+\sqrt{\frac{\Lambda}{3}}\right)\geq 0\;\;\;\;\;\;\; 
&\;\;\;\Longleftrightarrow\;\;\;&\;\;\;\;\;(H-{\Lambda}_{1})
(H-{\Lambda}_{2})\geq 0,\nonumber\\
\dot{H}\leq\frac{\Lambda}{3}-H^{2}\leq 0 
&\;\;\;\Longrightarrow\;\;\;&\;\;\;\;\; \dot{H}\leq 0\,.\nonumber
\eeqa
Essentially the equations involving $H$  
 are the same in both cases. The true difference concerns the equation 
for  $\dot H$; our condition  $(\dot H \leq 0)$ is more general than  
$\dot H \leq (\Lambda/3 - H^2)\leq 0$.
The hypothesys (\ref{iii}), when $\p \rightarrow $const is nothing else but 
$H^2 \geq \frac{\Lambda}{3}$ (in our unit $G_{eff} \rightarrow G_{N}$
if $F \rightarrow - \frac{1}{2}$); that is we recover the standard
case where $V=const$ is interpreted as the cosmological constant. 
By some algebra, it is easy to show that such a hypothesis is equivalent to 
\beq
\label{pippo}
\frac{1}{12F}\frac{{\dot \p}^2}{2V} \geq \left(\frac{F'}{F}\right)^{2} =
 \left(\frac{G'_{eff}}{G_{eff}}\right)^2 \,.
\eeq
That is the above hypothesis pose a constraint on the minimun value
(given by the relative (quadratic) variation of $G_{eff}$) of the (effective) 
ratio of the  kinetic energy  
and the potential energy of the scalar field.
Having shown that $a(t)$ behaves like de Sitter for large $t$,
we have to  see if it is possible to fix $\alp$ in order 
to recover  the effective
cosmological constant.
To this purpose, the Bianchi contracted identity for  matter is 
needed (it is important to stress that we have not used any Bianchi
contracted identity to find  the asymptotic behaviour of $a(t)$).
As usual, we get   $\rho_{m} = Da^{-3\gam}$
(we have used the state equation 
$p_{m} =(\gam -1)\rho_{m}$, with $1\leq \gam \leq 2$; $D$ is  the 
integration constant giving the matter content of universe). 
Introducing this expression for the matter in Eq.(\ref{7}),
for large $t$, we have 
\beq
(H - \Lambda_{1})(H + |\Lambda_{2}|) = 
\frac{D}{|F_{0}|}e^{-(3\gam\alp + \Sigma_{0})t}\,,
\eeq
being $3\gam \alp+\Sigma_{0} \geq 0$.
Then we get
\beq
(H - \Lambda_{1})(H + |\Lambda_{2}|)\rightarrow 0\,,
\eeq
i.e. $H \rightarrow \Lambda_{1}$.
The (effective) matter content, 
$\rho_{m}/6\f$,
tells us how $H$ is "distant" from the de Sitter behaviour given by
the cosmological constant $\Lambda_{1}$. In other words, we do not use 
the Bianchi identity for finding the type of expansion, 
we only use it to select (asymptotically) the specific value of 
what we call "cosmological constant". 
Of course we have that the universe undergoes
a de Sitter asymptotic expansion independently of any initial data.
Actually the effective cosmological constant that we have obtained
$\it {via}$ such a procedure will depend on the parameters connected to
the effective gravitational coupling "constant" and on those connected 
to the potential $V(\p)$.
Essentially, we have introduced the (effective) cosmological constant  
in a "pragmatic" way, through the 
(asymptotic) de Sitter behaviour for $a(t)$. In a certain sense,
the  approach in \bib{wald} is reversed: there, $\Lambda$ (constant) is
introduced {\it a--priori} 
and this leads, under certain hypotheses,  to a de Sitter expansion.
Here, the de Sitter expansion is recovered under completely different  
hypotheses, and this 
(together with the contracted Bianchi identity  for matter)
selects the effective cosmological constant.  
Moreover, we have 
obtained such a result without assuming to recover the standard
gravity (i.e. we do not need that $G_{eff}\rightarrow G_{N}$).
If we now consider also the KG equation, from the 
condition (\ref{ii}), 
we get, for large $t$, 
\beq
\label{12}
\frac{\dot\p^2}{F(\p)} =C_{1}(\Sigma_{0},\Sigma_{1}) =
- 2\Sigma_{0}\left(\sqrt{\left(\frac{\Sigma_{0}}{2}\right)^{2} + 
|\Sigma_{1}|}- \frac{3}{2}\Sigma_{0}\right)\,,
\eeq
that is $\dot{\p}^{2}/\f$ goes to a constant. Being $F(\p(t\gg 0))\leq 0$,
such a constant has to be negative: this request implies the following
relation between $\Sigma_{0}$ and $\Sigma_{1}$, which has to be satisfied 
for the sign compatibility:
\beq
\label{13}
|\Sigma_{1}| \geq 2\Sigma_{0}^{2}\,.
\eeq
By Eq.(\ref{13}) and condition (\ref{ii}), we get also
\beq
\label{14}
\frac{V}{6F} =  C_{2}(\Sigma_{0},\Sigma_{1}) = 
\left(\sqrt{\left(\frac{\Sigma_{0}}{2}\right)^{2} + |\Sigma_{1}|}\right)
\left(\frac{\Sigma_{0}}{6} - 
\sqrt{\left(\frac{\Sigma_{0}}{2}\right)^{2}+|\Sigma_{1}|}\right)\,.
\eeq
That is the potential has to be (asymptotically) nonnegative.
From the above  relations we see that in the case $\Sigma_{0} = 0$,
we get that only ${\displaystyle\frac{V}{6F}}$ is different from zero,
giving rise to the
expression ${\displaystyle\frac{V}{6F}}(t\gg 0) = -\Sigma_{1}^2$ which 
identifies,in this case, the cosmological (asymptotically) constant.
Finally from Eqs.(\ref{12}),(\ref{14}), we find:
\beq
\frac{\dot{\p}^{2}}{V} = C_{3}(\Sigma_{0},\Sigma_{1}) = 
\frac{2\Sigma_{0}\sqrt{\left(\frac{\Sigma_{0}}{2}\right)^2 + 
|\Sigma_{1}|}-
3\Sigma_{0}^{2}}{\Sigma_{0}\sqrt{\left(\frac{\Sigma_{0}}{2}\right)^2 
+|\Sigma_{1}|} 
-6\left(\frac{\Sigma_{0}}{2}\right)^2 - 6|\Sigma_{1}|}\,.
\eeq
We will show the relevance of these relations discussing some concrete 
examples at the end of the paper.
Let us consider now the other possibility connected to the case
$F(\p(t\gg 0))\leq 0$, that is $\dot F(\p(t \gg 0))\geq 0\,.$
In this case, $\Sigma_{0} \leq 0$ while everything else is the same as in 
the case  discussed above. 
In particular, the signs of the asymptotic values of 
$\Lambda_{1,2}$ are the same.
Referring to our previous analysis,
it is easy to show that now everything goes as in the Wald case
(as it is clear looking at (\ref{8''}),
that is it is possible to get the same two 
inequalities which are found in his proof).
It is interesting that the compatibility of all the  hypotheses  
that we have made with the KG equation gives rise again to Eq.(\ref{12}), 
but being $\Sigma_{0}\leq 0$, we get $\dot{\p}^2/\f\geq 0$ . 
Then the compatibility
between (\ref{ii}) and the KG equation implies, for large $t$, that the scalar
field has to go to a constant. In our units, $F\rightarrow -1/2$,
and $\Lambda \rightarrow \sqrt{V(t\gg 0)/3}$.

Finally, let us consider the case of asymptotically repulsive 
gravity, that is
\beq
F(\p(t\gg 0))\geq 0\,.
\eeq
Also here we have two subcases, $\dot{F}(\p(t \gg 0))\leq 0$ and
$\dot{F}(\p(t \gg 0))\geq 0$. 
As we have already stressed, even if this situation seems unphysical,
it gives a better understanding
of the non-necessary correlation between the (asymptotic) de Sitter
behaviour (i.e. between the {\it no--hair}) theorem and the recovering of  
standard gravity.
Of course, the condition on the reality of $\Lambda_{i}$ now has to be
carefully considered.
The most interesting case is $\dot{F}\leq 0$. Here, we have 
two (asymptotic)
positive cosmological constants, that is  
\beq
\Lambda_{eff\,1,2}\rightarrow \Lambda_{1,2}\geq 0\,, 
\;\;\;\;\mbox{with}\;\;\;\;
\Lambda_{1}\geq \Lambda_{2}\,.
\eeq
Being $-\rho_{m}/6F\leq 0$,
we have $\Lambda_{1}\leq H\leq \Lambda_{2}$. Then, it is crucial 
to know the sign of $\dot H$: if $\dot H \geq 0$ the effective $\Lambda$
is given by the $max\,(\Lambda_{1},\Lambda_{2})$; viceversa,
if $\dot H \leq 0$,  $\Lambda$ is given by the 
minimun between them. 
We will discuss an example of this last situation.

\section{\normalsize \bf Examples}
Now we  present some realizations of the above discussion.
First of all, we have that the field Lagrangian 
(density), giving rise to the action (\ref{1}),
becomes in the FRW (flat) case:

\beq
\label{pointlike}
{\cal L}={\l}_{\p}+Da^{3(1-\gam)}\,,
\eeq
where
\beq
{\l}_{\p}=
6a\dot{a}^{2}\f+6\dot{a}\dot{\p}a^{2}\f{\p}
+a^{3}\left(\frac{1}{2}\dot{\p}^{2}-\v \right)\,.
\eeq
We  restrict our analysis to a dust--dominated universe
 $(\gam=1)$,that is to the case ${\l}={\l}_{\p}+D,$
since we are interested in asymptotic regimes.
\begin{enumerate}
\item 
The simplest example is given by $\p=const$, $\f=-1/2$ and
$\v=\Lambda$, that is the standard de Sitter case.
In this case we have $\Sigma_{0}=0$ and $\Sigma_{1}=-\Lambda/3$.
\item
Let us consider an {\it a--priori} generic nonminimal coupling $F(\p)$
and the potential $\v= \Lambda$. Using the  N\"other Symmetry Approach
\bib{cimento},\bib{nmc},
we get $\f=\frac{1}{12}\p^{2}+F_{0}'\p+F_{0}$, 
where $F_{0}'$ and $F_{0}$ are two
generic parameters. We have
already discussed such a case in \bib{cimento},\bib{nmc}.
From the relation  between the asymptotic behaviour 
of the potential and the coupling(relation that we have found using the 
compatibility  between the hypotheses we have done and the KG eq.)we 
see that the coupling has to go,asymptotically, to a constant.
Anyway the general solution, for $\gam=1$, is
\beq
a(t)=\left[c_{1}e^{\lambda t}
+c_{2}e^{-\lambda t}\right]^{1/2}\,,
\eeq
and
\beq
\p(t)=\frac{{\cal J}_{0}}{\sqrt{c_{1}e^{\lambda t}+c_{2}e^{-\lambda t}}}
{\cal K}+\frac{c_{3}}{\sqrt{c_{1}e^{\lambda t}+c_{2}e^{-\lambda t}}}-6F_{0}'\,,
\eeq
where $c_{1}$, $c_{2}$ and $c_{3}$ are the three integration constants and 
$\lambda =\sqrt{-2\Lambda/3{\cal H}}$.
${\cal J}_{0}$ is a constant of motion, 
 ${\cal H}=F_{0}-3{F'}_{0}^{2}\leq 0$ is the
$\p$--part of  Hessian determinant of $\cal L$,which depends only on the 
parameters connected to the function describing the coupling, and 
\beq
{\cal K}=\int \frac{dt}{\sqrt{c_{1}e^{\lambda t}+c_{2}e^{-\lambda t}}}\;,
\eeq
is an elliptical integral of first kind.
In this case, the effective cosmological constant 
is asymptotically given by 
\beq
-\frac{\Lambda}{6F(\p(t\gg 0))}=\frac{V(\p(t\gg 0))}{3{|\cal H|}}\,, 
\;\;\;\;\;\;\;\;\;
|\Sigma_{1}|=\frac{\Lambda}{6{|\cal H|}}\,.
\eeq
since $\Sigma_{0}\rightarrow 0$ $(\p \rightarrow -6F'_{0})$.
Then, for $t\rightarrow \infty$, we have
$ a(t)\sim e^{\frac{\lambda}{2} t}\,.$
In this case, the  conditions (\ref{i}) and (\ref{ii}) hold,
and the standard Einstein gravity
$(G_{eff}\rightarrow G_{N})$ is restored. Of course $\lambda=
2\sqrt{{\displaystyle\frac{V(\p(t\gg 0))}{3 |\cal{H}|}}}$.
\item
In the case 
$\f=k_{0}\p^{2},\;\;\;\;\v=\lambda\p^{2},\;\;\;\;\gam=1\,,$
where $k_{0}<0$ and $\lambda > 0$ are free parameters,
the de Sitter regime is recovered even if solutions do not converge toward 
standard gravity.
The coupling $\f$ is always negative, whereas 
$\v$ is always positive and $\dot F(\p(t \gg 0))<0$.
Infact the general solutions 
are \bib{cimento},\bib{nmc}
\beqa
a(t)&=&\left[
c_{1}e^{{\Lambda_{0}} t}+c_{2}e^{-{\Lambda_{0}} t}\right]\times
\nonumber \\
    & & \exp\left\{ -\frac{2}{3}\left[
c_{3}\arctan \sqrt{\frac{c_{1}}{c_{2}}} e^{{\Lambda_{0}} t}
+c_{4}\ln (c_{1}e^{{\Lambda_{0}} t}+c_{2}e^{-{\Lambda_{0}} t})\right]\right\},
\label{014}
\eeqa
which is clearly de Sitter for $t\gg 0$, and
\beq
\label{015}
\p (t)=
\frac{\exp\left[c_{3}\arctan \sqrt{\frac{c_{1}}{c_{2}}} e^{{\Lambda_{0}} t}
+c_{4}\ln (c_{1}e^{{\Lambda_{0}} t}+c_{2}e^{-{\Lambda_{0}} t})
\right]}{c_{1}e^{{\Lambda_{0}} t}+c_{2}e^{-{\Lambda_{0}} t}}.
\eeq
where
\beq
\Lambda_{0}=\sqrt{\frac{2\lambda \xi_{2}}{\xi_{1}(\xi_{1}-\xi_{2})}},\;\;
c_{3}=\frac{{\cal F}_{0}\sqrt{c_{1}c_{2}}}{\xi_{2}\Lambda_{0}},
\;\;c_{4}=\frac{\xi_{1}}{\xi_{2}}\,,
\;\;\xi_{1}=1-12k_{0},\;\;\xi_{2}=1-\frac{32}{3}k_{0}\,.
\eeq
The constants $c_{1}$, $c_{2}$, $c_{3}$  are  the initial data 
and ${\cal F}_{0}$
is a constant of motion related to the existence of the N\"other 
symmetry \bib{cimento},\bib{nmc}. 
We get asymptotically
\beq
\Sigma_{0}=\sqrt{\frac{-32\lambda k_{0}}{(1-12k_{0})(3-32k_{0})}}\,,\;\;
\Sigma_{1}=\frac{\lambda\left(128k_{0}^{2}-24k_{0}+
1\right)}{2k_{0}(3-32k_{0})(1-12k_{0})}\,.
\eeq
In this case, it is always $\Sigma_{0}> 0$.
Finally
\beq
\Lambda_{eff,\,1}=-\frac{\dot{F}}{2F}+
\sqrt{\left(\frac{\dot{F}}{2F}\right)^{2}-\frac{\rho_{\p}}{6F}}\longrightarrow
\Lambda=
\sqrt{\frac{\lambda(1-8k_{0})^{2}}{2k_{0}(12k_{0}-1)(3-32k_{0})}}\ > 0\,,
\eeq
which is exactly the constant that appears in the asymptotic 
behaviour of the
scale factor $a(t)\sim \exp(\Lambda t)$, 
i.e.  the effective 
cosmological constant.
It is relevant to stress that, we have
\beq
F(\p(t\gg 0))\rightarrow  
k_{0}
\exp\left[2\Lambda\left(\frac{4|k_{0}|}{3+32|k_{0}|}t\right)\right]\ < 0\,,
\eeq
and $F(\p(t))$ diverges. We do not recover asymptotically the standard 
$G_{N}$. Actually we have  (plus infinity) asymptotic gravitational
freedom \bib{freedom}:nevertheless we have a de Sitter behaviour at infinity 
for $a(t)$.
Furthermore, the condition (\ref{ii}) is always satisfied.
\item
Another interesting case is
$\f = k_{0}\p^2,\,\,\,\v=\lambda \p^2,\,\,\,\gam=1,$
with $k_{0}>0$ (precisely  $1/12 <k_{0}< 3/32$).
The solutions are essentially the same as in the case $k_{0}<0$, except
we have to change "arctan" with "arctanh".
The asymptotic behviours of  $a(t)$ and $\p(t)$ are:
\beq
a(t) \sim e^{\Lambda_{0} (1-8k_{0})/(3-32k_{0})t},\,\,\,\, 
\p(t) \sim e^{\Lambda_{0} (4k_{0})/(32 k_{0}-3)t}, 
\eeq
Now  we have  $32 k_{0}-3<0$ and then $\p(t)$ is a decreasing function 
of time, which implies  $\dot F(\p(t \gg t)) \leq 0$ and $\Sigma_{0}\leq 0$.
We see that 
$\dot H \leq 0$ and then the effective cosmological constant is given by the
$min\,\,(\Lambda_{1},\Lambda_{2})$.
By some algebra, it is possible to verify that the true cosmological constant
is $\Lambda_{2}$ (which is always less than $\Lambda_{1}$).
This example is useful to stress that the de Sitter asymptotic behaviour,
connected with the presence of a cosmological constant, 
is independent of the sign of 
gravitational coupling.
\item
The last   case  we discuss is $F=-1/2$, 
$\v = V_{0}(Ae^{2\lambda\p}+Be^{2\lambda\p})^2$,
where $A, B$ and $V_{0}$ are constants,and $\lambda$ is given
in terms of $G_{N}$ (see \bib{minimal} for details).  
We are in a situation similar to that discussed in our second example.
Using the asymptotic relation between the coupling and the 
potential,also in this case the potential has to go (asymptotically)to 
a constant.
Anyway,we have used our N\"other symmetry approach for 
solving exactly the model.
Asymptotically,using the behaviour of the exact solutions  we find
\beq
a(t) \sim e^{\Lambda t}\,,\,\,\,\,\,
\p \sim \it{const},\,
\eeq
where $\Lambda$ at the exponent of $a(t)$ is:
\beq
\Lambda =\sqrt{\frac{4|AB|}{3}}\,.
\eeq
If we compute the effective (positive) $\Lambda$ from the definition (\ref{9}),
we find 
\beq
\sqrt{-\frac{\v}{6F}} = \sqrt{ \frac{4|AB|}{3}}\;;
\eeq
i.e. the same quantity as given by the asymptotic behaviour of the scale 
factor.
Of course the standard matter has no role in this asymptotic regime.

We conclude the discussion of these examples stressing, again, that
it appears clear that
the (asymptotic) cosmological constant,as introduced in our
approach,is a function of the parameters 
appearing into the two functions $F(\p)$, $V(\p)$.

\end{enumerate}

\section{\normalsize \bf  Conclusions}

We have discussed the cosmic no--hair theorem in the framework
of nonminimally coupled scalar--tensor theories. We have 
introduced a time dependent cosmological "constant" not using the "geometrical
side" of such theories (i.e. $\Lambda g\dmunu$,as usual) but the "scalar
side". That is the effective cosmological "constant" has been reconstructed
by  $\dot{G}_{eff}/G_{eff}$ and by $\rho_{\p}/6\f$.
Actually $\Lambda_{eff}$ has been introduced only in the case 
of homogeneous--isotropic flat cosmologies but it is not difficult to extend
the above considerations to any Bianchi model (except Bianchi IX).
The  way we have followed to recostruct the no--hair theorem is 
opposite of that  usually adopted: instead of introducing {\it by hand}
a cosmological constant and then searching for the conditions for an 
asymptotic de Sitter behaviour,
we  find the conditions to get such an  asymptotic behaviour,
and then we  define an effective cosmological "constant"
(actually function of time), which becomes a (true) constant for $t\gg 0$.
Of course, the time behaviour of  $\Lambda_{eff}$ can be  of any
type with respect to the asymptotic constant value \bib{matarrese}.  
Under the hypotheses we  used, the de Sitter asymptotic regime is 
obtained and this is not necessarely connected with  
recovering the standard Einstein
gravity (which is restored, in our units, for the value 
$\f_{\infty}=-1/2$ of the coupling). 
In other words, the cosmic no--hair theorem
 holds even if we are not in the Einstein regime (it is not even necessary
that the right (attractive gravity) sign of the coupling is recovered). 
Furthermore, the role of the Bianchi contracted
identity for the (standard) matter is to fix (only) 
the specific value of $\Lambda$,
not the kind of the (de Sitter) asymptotic behaviour of $a(t)$.
It is interesting to stress that, by this mechanism, the "amount of $\Lambda$"
is strictly related to the matter content of the universe.
This is worthwhile in connection to the $\Omega$ problem
since it seems that cold dark matter models, with non trivial amount
of cosmological constant, have to be taken into serious consideration for
large scale structure formation \bib{starobinsky}. 
In conclusion, we want to make two final remarks. The first concerns 
an important question which we have only
mentioned. The way we have followed to introduce the (effective)
cosmological "constant" seems to confine its meaning  only to the cosmological
arena. In the  standard way used to define  such a
quantity, the problem does not exists since it is a true constant of the 
theory and then it is defined independently of any cosmological scenario.
We believe that this question can be solved stressing that cosmology 
has to be taken into account in any other specific physical situation
in relativity. Then  the effective time--dependent
cosmological constant we have introduced gets a role 
of the same kind of the standard $\Lambda$.
From this point of view,  the question we 
are discussing can be answered still using the (standard) way to
define the  cosmological constant, i.e.(the cosmological)  $T_{0 0}$. This is
what we actually have done  and what we believe to be
the ingredient to use for understanding the role of  
(effective) cosmological "constant" also in different contexts than cosmology.
Finally, in our construction of $\Lambda$, there is 
a contribution given by the (relative) time variation of the 
effective gravitational coupling: this implies that it would be possible
to compute it, for example, $\it via$ the density contrast
parameter. This will be our next step in this kind of research. 

\vspace{2. mm}
{\bf ACKNOWLEDGEMENT}. We have the pleasure to acknoledgement helpful
discussions with  A.Marino,G Marmo.

\begin{centerline}
{\bf REFERENCES}
\end{centerline}
\begin{enumerate}
\item \label{wald}
R.M.Wald, \pr {\bf D28} (1983) 2118.
\item\label{bergmann}
P.G. Bergmann, \ijtp  {\bf 1}  (1968) 25.
\item\label{brans}
D.W. Sciama, {\it Mon. Not. R. Astrom. Soc.} {\bf 113} (1953) 34\\
C. Brans  and R.H. Dicke,  {\it Phys. Rev} {\bf 124} (1961) 925
\item\label{cimento}
S. Capozziello, R. de Ritis, C. Rubano, and P. Scudellaro,
to appear in {\it La Rivista del Nuovo Cimento} (1996).
\item \label{hoyle}
F.Hoyle and J.V.Narlikar, {\it Proc. R. Soc.}, {\bf A 273} (1963) 1.
\item\label{cotsakis}
S. Cotsakis and G. Flessas, \pl {\bf B 319} (1993) 69;
A.B.Burd and J.D. Barrow,\np {\bf B308} (1988) 929;
J.Yokogawa and K.Maeda, \pl {\bf B207} (1988) 31;
J.D. Barrow and G. G\"otz, \pl {\bf B 231} (1989) 228.
\item\label{Pimentel}
L.O.Pimentel,J.Stein--Schabes,\pl {\bf B 216} (1989) 27.
\item \label{nmc}
S. Capozziello and R. de Ritis, \pl {\bf A 177} (1993) 1;
S. Capozziello, R. de Ritis, and P. Scudellaro,  \ijmp {\bf D 2} (1993) 463;
S. Capozziello and R. de Ritis, \cqg {\bf 11} (1994) 107;
S. Capozziello, R. de Ritis, and P. Scudellaro, \pl {\bf A 188} (1994) 130.
\item\label{will}
C.M. Will {\it Theory and experiments in gravitational physics},
Cambridge Univ. Press, Cambridge (1993).
\item\label{minimal}
R. de Ritis, G. Marmo, G. Platania, C. Rubano, P. Scudellaro, and C. 
Stornaiolo, \pr {\bf D 42} (1991) 1990.
\item\label{freedom}
S. Capozziello and R. de Ritis,\pl {\bf A 208} (1995) 181.
\item\label{matarrese}
M. Bruni, S. Matarrese, O. Pantano, \prl {\bf 74} (1995) 1916.
\item\label{starobinsky}
A.A. Starobinsky, in {\it Cosmoparticle Physics} 1, eds. M.Yu. Khlopov 
{\it et al.}, Edition Frontiers (1996).
\end{enumerate}

\vfill

\end{document}